\documentclass{aa}
\usepackage{natbib}
\usepackage{xspace}
\usepackage{amssymb}
\usepackage{amsmath}
\usepackage{graphicx}

\def\aap{A\&A}

\def\apj{ApJ}

\begin{document}
\title{A representative particle approach to coagulation and fragmentation of
dust aggregates and fluid droplets}
\titlerunning{A representative particle approach}
\authorrunning{Zsom \& Dullemond}
\author{A.~Zsom$^1$ and C.P.~Dullemond$^{1}$}
\institute{(1) Max-Planck-Institut f\"ur Astronomie, K\"onigstuhl 17,
  D-69117 Heidelberg, Germany.\\
  \email{zsom@mpia.de, dullemon@mpia.de}
}
\date{\today}

\abstract{
  {\em Context:} There is increasing need for good algorithms for modeling
  the aggregation and fragmentation of solid particles (dust grains, dust
  aggregates, boulders) in various astrophysical settings, including
  protoplanetary disks, planetary- and sub-stellar atmospheres and dense
  molecular cloud cores. Here we describe a new algorithm that combines
  advantages of various standard methods into one.
\\
{\em Aims:} The aim is to develop a method that 1) can solve for aggregation
and fragmentation, 2) can easily include the effect and evolution of grain
properties such as compactness, composition, etc., and 3) can be built as a
coagulation/fragmentation module into a hydrodynamics simulation where it
3a) allows for non-`thermalized' non-local motions of particles
(e.g.~movement of particles in turbulent flows with stopping time larger
than eddy turn-over time) and 3b) focuses computational effort there where
most of the mass is.
\\
{\em Methods:} We develop a Monte-Carlo method in which we follow the
``life'' of a limited number of representative particles. Each of these
particles is associated with a certain fraction of the total dust mass and
thereby represents a large number of true particles which all are assumed to
have the same properties as their representative particle. Under the
assumption that the total number of true particles vastly exceeds the number
of representative particles, the chance of a representative particle
colliding with another representative particle is negligibly small, and we
therefore ignore this possibility. This now makes it possible to employ a statistical approach to the evolution of the representative
particles, which is the core of our Monte Carlo method.
\\
{\em Results:} The method reproduces the known analytic solutions of
simplified coagulation kernels, and compares well to numerical results for
Brownian motion using other methods. For reasonably well-behaved kernels it
produces good results even for moderate number of swarms.}

\maketitle

\begin{keywords}
accretion, accretion disks -- circumstellar matter 
-- stars: formation, pre-main-sequence -- infrared: stars 
\end{keywords}

\section{Introduction}
Dust particle aggregation is a very common process in various astrophysical
settings. In protoplanetary disks the aggregation of dust particles forms
the very initial step of planet formation (see e.g.\ Dominik et
al.~\citeyear{dominikblum:2007}). It also modifies the optical properties of
the disk, and it has influence on the chemistry and free electron abundance
in a disk (Sano et al.~\citeyear{sanomirama:2000}; Semenov et
al.~\citeyear{semenovwiebe:2004}; Ilgner \& Nelson
\citeyear{ilgnernelson:2006a}). The appearance and evolution of a
protoplanetary disk is therefore critically affected by the dust aggregation
process. In sub-stellar and planetary atmospheres the aggregation of dust
particles and the coagulation of fluid droplets can affect the
structure of cloud layers. It can therefore strongly affect the spectrum of
these objects and influence the local conditions within these
atmospheres. The process of aggregation/coagulation and the reverse process
of fragmentation or cratering are therefore important processes to
understand, but at the same time they are extremely complex.

Traditional methods solve the Smoluchowski equation for the particle mass
distribution function $f(m)$, where $f(m)$ is defined such that $f(m)dm$
denotes the number of particles per cubic centimeter with masses in the
interval $[m,m+dm]$. This kind of method has been used in many papers on
dust coagulation before (e.g.\ Nakagawa et
al.~\citeyear{nakanakahayashi:1981}; Weidenschilling
\citeyear{weidenschilling:1984,weid1997}; Schmitt et
al.~\citeyear{schmitthenningmucha:1997}; Suttner \& Yorke
\citeyear{suttneryorke:1999}; Tanaka et al.~\citeyear{tanakahimemoida:2005};
Dullemond \& Dominik \citeyear{duldom:2005}; Nomura \& Nakagawa
\citeyear{nomuranaka:2006}). Methods of this kind are efficient, but have
many known problems. First of all a coarse sampling of the particle mass
leads to systematic errors such as the acceleration of growth (Ohtsuki et
al.~\citeyear{ohtsuki:1990}). High resolution is therefore required, which
may make certain problems computationally expensive. Moreover, if one wishes
to include additional properties of a particle, such as porosity, charge,
composition etc, then each of these properties adds another dimension to the
problem. If each of these dimensions is sampled properly, this can
quickly make the problem prohibitively computationally expensive. Finally,
the traditional methods are less well suited for modeling stochastic
behavior of particles unless this stochastic behavior can be treated in an
averaged way. For instance, in protoplanetary disks if the stopping time of
a particle is roughly equal to the turbulent eddy turn-over time, then the
velocity of a particle with respect to the gas is stochastic: at the same
location there can exist two particles with identical properties but which
happen to have different velocities because they entered the eddy from
different directions (see e.g.\ the simulations by Johansen et
al.~\citeyear{johansen:2006}).

To circumvent problems of this kind Ormel et
al.~(\citeyear{ormelmonte:2007}) have presented a Monte Carlo approach to
coagulation. In this approach the particles are treated as computational
particles in a volume which is representative of a much larger volume. The simulation follows the life of $N$ particles as they collide
and stick or fragment. The collision rates among these particles are
computed, and by use of random numbers it is then determined which particle
collides with which.  The outcome of the collision is then determined
depending on the properties of the two colliding particles and their
relative impact velocity. This method, under ideal conditions, provides the true simulation of
the process, except that random numbers are used in combination with
collision rates to determine the next collision event. This method has many
advantages over the tradiational methods. It is nearly trivial to add any
number of particle properties to each particle. There is less worry of
systematic errors because it is so close to a true simulation of the system,
and it is easy to implement.  A disadvantage is that upon coagulation the
number of computational particles goes down as the particles
coagulate. Ormel et al.~solve this problem by enlarging the volume of the
simulation and hence add new particles, but this means that the method is
not very well suited for modeling coagulation within a spatially resolved
setting such as a hydrodynamic simulation or a model of a protoplanetary
disk.

It is the purpose of the present paper to present an alternative Monte Carlo
method which can quite naturally deal with extremely large numbers of
particles, which keeps the number of computational particles constant
throughout the simulation and which can be used in spatially resolved
models.

\section{The method}
\subsection{Fundamentals of the method}\label{subsec-fundamentals}
The fundamental principle underlying the method we present here is to follow
the behavior of a limited number of {\em representative particles} whose
behavior is assumed to be a good representation of all particles.  In this
approach the number of physical particles $N$ can be arbitrarily large. In
fact it {\em should} be very much larger than the number of representative
particles $n$, so that the chance that one representative particle collides
with another representative particle is negligible compared to the
collisions between a representative particle and a non-representative
particle. In other words, if $N\gg n$, we only need to consider collisions
between a representative particle and a non-representative particle. The
number of collisions among representative particles is too small to be
significant, and the collisions among non-representative particles are
not considered because we focus only on the behavior of the representative
particles.

Suppose we have a cloud of dust with $N=10^{20}$ physical particles, with a
specific size distribution, for instance, MRN (Mathis, Rumpl \& Nordsieck
\citeyear{mrn:1977}). Let the total mass of all these particles together be
$M_{\mathrm{tot}}$ and the volume be $V$. We randomly pick $n$ particles out
of this pool, where $n$ is a number that can be handled by a computer, for
instance, $n=1000$. Each representative particle $i$ has its own mass $m_i$
and possibly other properties such as porosity $p_i$ or charge $c_i$
assigned to it. We now follow the life of each of these $n=1000$
particles. To know if representative particle $i=20$ collides with some
other object, we need to know the distribution function of all {\em
  physical} particles with which it can collide. However, in the computer we
only have information about the $n$ representative particles. We therefore
have to make the assumption that the distribution function set up by the $n$
representative particles is representative of that of the $N$ physical
particles. We therefore assume that there exist
\begin{equation}
n_k=\frac{M_{\mathrm{tot}}}{nm_kV}
\end{equation}
physical particles per cubic cm with mass $m_k$, porosity $p_k$, charge
$c_k$ etc., and the same for each value of $k$, including $k=i$. In this
way, by assumption, we know the distribution of the $N$ physical particles
from our limited set of $n$ representative particles. One could say that
each representative particle represents a swarm of $M_{\mathrm{tot}}/nm_i$
physical particles with identical properties as the representative one. One could also say that the
true distribution of $N$ particles is, by assumption, that of the $n$
representative ones.  The rate of collisions that representative particle
$i$ has with a physical particle with mass $m_k$ etc.\ is then:
\begin{equation}
r_i(k) = n_k\sigma_{ik}\Delta v_{ik}=\frac{M_{\mathrm{tot}}}{nm_kV}\sigma_{ik}\Delta v_{ik}
\end{equation}
where $\sigma_{ik}$ is the cross-section for the collision between particles
with properties $i$ and $k$, and $\Delta v_{ik}$ is the average relative
velocity between these particles. The total rate of collisions that
representative particle $i$ has with any particle is then:
\begin{equation}
r_i = \sum_{k}r_i(k)
\end{equation}
and the total rate of collisions of any representative particle is
\begin{equation}
r = \sum_i r_i
\end{equation}

The time-evolution of the system is now done as follows. Let $t_0$ be the
current time. We now randomly choose a time step $\delta t$ according to:
\begin{equation}\label{eq-delta-t}
\delta t = -\frac{1}{r}\log(\mathrm{ran(seed)})
\end{equation}
where ran(seed) is a random number uniformly distributed between 0 and 1.
This means that a collision event happens with one of the representative
particles at time $t=t_0+\delta t$. The chance $P(i)$ that the event happens
to representative particle $i$ is:
\begin{equation}
P(i) = \frac{r_i}{r}
\end{equation}
So we can choose, using again a random number, which representative particle
$i$ has undergone the collision event. We now need to determine with what kind
of physical particle it has collided. Since the distribution of physical
particles mirrors that of the representative ones, we can write that the
chance this particle has collided with a physical particle with
properties $k$ is:
\begin{equation}
P(k|i) = \frac{r_i(k)}{r_i}
\end{equation}
With another random number we can thus determine which $k$ is involved in
the collision. Note that $k$ can be $i$ as well, i.e.\ the representative
particle can collide with a physical particle with the same properties, or
in other words: a representative particle can collide with a particle of
its own swarm of physical particles.

Now that we know what kind of collision has happened, we need to determine
the outcome of the collision. The most fundamental part of our algorithm is
the fact that only representative particle $i$ will change its properties in
this collision. Physical particle $k$ would in principle also do so (or in
fact becomes part of the new representative particle), but since we do not
follow the evolution of the physical particles, the collision will only
modify the properties of representative particle $i$. By assumption
this will then automatically also change the properties of all physical
particles associated with representative particle $i$. Statistically,
the fact that the particles $k$ are not modified is ``corrected for'' by
the fact that at some point later the {\em representative} particle $k$ will
have a collision with {\em physical} particle $i$, in which case the 
properties of the $k$ particles will be modified and not those of $i$. This
then (at least in a statistical sense) restores the ``symmetry'' of the
interactions between $i$ and $k$. If the collision
leads to sticking, then the resulting particle will have mass
$m=m_i+m_k$. This means that representative particle $i$ will from now on
have mass $m_i\leftarrow m_i+m_k$. Representative particle $k$ is left
unaffected as it is not involved in the collision. The interesting thing is
now that, because by assumption the representative particle distribution
mirrors the real particle distribution, the swarm of physical particles
belonging to the modified representative particle $i$ now contains fewer
physical particles, because the total dust mass $M=M_{\mathrm{tot}}/n$ of
the swarm remains constant.

If a collision results in particle fragmentation, then the outcome of the
collision is a distribution function of debris particles. This distribution
function can be written as a function $f_d(m)$ of debris particle mass, such
that
\begin{equation}
\int_0^{\infty}mf_d(m)dm=m_i+m_k
\end{equation}
and the function $f_d(m)$ has to be determined by laboratory experiments or
detailed computer simulations of individual particle collisions (see Dominik
et al.\ \citeyear{dominikblum:2007} for a review). The new value of $m_i$
for the representative particle is now randomly chosen according to this
distribution function by solving the equation
\begin{equation}\label{eq-solve-new-debris-mass}
\int_0^{\bar m}mf_d(m)dm = \mathrm{ran(seed)}(m_i+m_k)
\end{equation}
for $\bar m$ and assigning $m_i\leftarrow \bar m$. In other words: we
randomly choose a particle mass from the debris mass distribution function,
i.e.\ the choice is weighed by fragment mass, not by fragment particle
number. This can be understood by assuming that the true representative
particle before the collision is in fact just a monomer inside a larger
aggregate. When this aggregate breaks apart into for instance one big and
one small fragment it is more likely that this representative monomer
resides in the bigger chunk than in the smaller one.

After a fragmenting collision the $m_i$ will generally be smaller than
before the collision. This means that the number of physical particles
belonging to representative particle $i$ increases accordingly. Note that
although the collision has perhaps produced millions of debris particles out
of two colliding objects, our method only picks one of these debris
particles as the new representative particle and forgets all the
rest. Clearly if only one such destructive collision happens, the
representative particle is not a good representation of this entire cloud of
debris products. But if hundreds such collisions happen, and are treated in
the way described here, then the statistical nature of
Eq.~(\ref{eq-solve-new-debris-mass}) ensures that the debris products are
well represented by the representative particles.

The relative velocity $\Delta v$ can be taken to be the average relative
velocity in case of random motions, or a systematic relative velocity in
case of systematic drift. For instance, for Brownian motion there will be an
average relative velocity depending on the masses of both particles, but
differential sedimentation in a protoplanetary disk or planetary atmosphere
generates a systematic relative velocity. Also, for the Brownian motion or
turbulent relative velocity one can, instead of using an average relative
velocity, choose randomly from the full distribution of possible relative
velocities if this is known. This would allow a consistent treatment of
fluctuations of the relative velocities which could under some circumstances
become important (see e.g.\ Kostinski \& Shaw \citeyear{kostinskishaw:05}).

\subsection{Computer implementation of the method}
We implemented this method in the following way. For each of the
representative particles we store the mass $m_i$ and all other properties
such as porosity, charge, composition etc. Before the start of the Monte
Carlo procedure we compute the full collision rate matrix $r_i(k)$, and we
compute the $r_i$ as well as $r$. For these collision pairs $(i,k)$ we now have to determine the
cross section of particles as well as their systematic relative velocity, such as different drift speeds, and the random relative velocity, such as Brownian or turbulent motion. The random motions
can be determined with a random number from the relative velocity
probability distribution function if that is known. If that is not known in
sufficient detail, one can also take it to be the average relative velocity,
for which more often analytic formulae exist in the literature.

We determine beforehand at which times
$t_{\mathrm{sav},n}$ we want to write the resulting $m_i$ and other
parameters to a file. The simulation is now done in a subroutine with a
do-while loop. We then determine $\delta t$ using a random number (see
Eq.~\ref{eq-delta-t}), and check if $t+\delta t< t_{\mathrm{sav},n}$, where
$t_{\mathrm{sav},n}$ is the next time when the results will have to be
stored. If $t+\delta t< t_{\mathrm{sav},n}$, then a collision event occured
before $t_{\mathrm{sav},n}$. We will handle this event according to a
procedure described below, we set $t\leftarrow t+\delta t$ and then return
to the point where a new $\delta t$ is randomly determined. If, on the other
hand, $t+\delta t\ge t_{\mathrm{sav},n}$ then we stop the procedure, return
to the main program and set $t\leftarrow t_{\mathrm{sav},n}$. The main
program can then write data to file and re-call the subroutine to a time
$t_{\mathrm{sav},n+1}$ or stop the simulation altogether. Note that when the
subroutine is called again for a next time interval, it does not need to
know the time of the previously randomly determined event which exceeded
$t_{\mathrm{sav},n}$. Of course, one could memorize this time and take that
time as the time of the next event in the next time interval, but since the
events follow a Poisson distribution, we do not need to know what
happened before $t_{\mathrm{sav},n}$ to randomly determine the new time
$t+\delta t$ of the next event. 

Now let us turn to what happens if a collision event occurs, i.e.\ occurs
between time $t$ and $t+\delta t$. We then first determine which
representative particle $i$ is hit, which is done by generating a random
number and choosing from the probability distribution of collision rates, as
described in Section \ref{subsec-fundamentals}. Similarly we determine the
non-representative particle with which it collides, or in other words: we
determine the index $k$ of the ``swarm'' in which this non-representative
particle resides. Finally, we must determine the impact parameter of the collision, or assume some average
impact parameter.

Now we employ a model for the outcome of the collision. This is the
collision subroutine of our Monte Carlo method. It is here where the results of
laboratory experiments come in, and the translation of such experiments into
a coagulation kernel is a major challenge which we do not cover here.  The
collision model must be a quick formula or subroutine that roughly
represents the outcomes of the detailed laboratory collision experiments or
detailed numerical collision models. It will give a probability function
$f_d$ for the outcoming particle masses and properties. From this
distribution function we pick {\em one} particle, and from this point on our
representative particle $i$ will attain this mass and these properties. The
collision partner $k$ will not change, because it is a non-represetative
particle from that swarm that was involved in the collision, and we do not
follow the life of the non-representative particles. We therefore ignore any
changes to that particle.

We now must update $r_j(l)$ for all $l$ with fixed $j=i$ and for all $j$
with fixed $l=k$: we update a row and a column in the $r_j(l)$
matrix. Having done this, we must also update $r_j$ for all $j$. This would
be an $n^2$ process, which is slow. But in updating $r_j(l)$ we know the
difference between the previous and the new value, and we can simply add
this difference to $r_j$ for each $j$. Only for $j=i$ we must recompute the
full $r_j$ again, because there all elements of that row have been modified.
Using this procedure we assure that we limit the computational effort to
only the required updates.

\subsection{Acceleration of the algorithm for wide size distributions}
\label{subsec-acceleration}
One of the main drawbacks of the basic algorithm described above is that it
can be very slow for wide size distributions. Consider a swarm of micron
sized dust grains that are motionless and hence do not coagulate among each
other. Then a swarm of meter sized boulders moves through the dust swarm at
a given speed, sweeping up the dust. Let us assume that also the boulders
are not colliding among each other. The only mode of growth is the
meter-sized boulders sweeping up the micron sized dust. For the boulder to
grow a factor of 2 in mass it will have to sweep up $10^{18}$ micron sized
dust particles.  Each impact is important for the growth of the boulder, but
one needs $10^{18}$ such hits to grow the boulder a factor of 2 in mass. The
problem with the basic algorithm described above is that it is forced to
explicitly model each one of these $10^{18}$ impacts. This is obviously
prohibitively expensive.

The solution to this problem lies in grouping collisions into one. Each
impact of a dust grain on a boulder only increases the boulder mass by a
minuscule fraction. For the growth of the boulder it would also be fine to
lower the chance of an impact by $10^{16}$, but {\em if} it happens, then
$10^{16}$ particles impact onto the boulder at once. Statistically this
should give the same growth curve, and it accelerates the method by a huge
factor. However, it introduces a fine-tuning parameter. We must specify the
minimum increase of mass for coagulation ($dm_{max}$). If we set $dm_{max}$ to, for instance,
10\%, then we may expect that the outcome also has errors of the order
  of 10\%. This error arises because by increasing the mass of the
  bigger body in steps of 10\%, we ignore the fact that the mass at some
  time should in fact be somewhere in between, which cannot be resolved with
  this method. This is, however, not a cumulative error. While the mass of
  the bigger body may sometimes be too low compared to the real one, it
  equally probably can be too large. On average, by the Poisson nature of
  the collision events, this averages out. But it is clear that the smaller
  we take this number, the more accurate it becomes -- but also the slower
  the method becomes. It is therefore always a delicate matter to choose
  this parameter, but for problems with a large width of the size
  distribution this acceleration is of vital importance for the usability of
  the method.

\subsection{Including additional particle properties}
We mentioned briefly the possibility of adding more particle properties to
each representative particle. This is very easy to do, and it is one of the
main advantages of a Monte Carlo method over methods that directly solve the
integral equations for coagulation. One of the main properties of interest
to planet formation is porosity or fractal structure of the aggregate. Two
aggregates with the same mass can have vastly different behavior upon a
collision if they have different compactness. A fluffy aggregate may break
apart already at low impact velocities while a compact aggregate may simply
bounce. Upon collisions these properties may in fact also change. Ormel et
al.~(\citeyear{ormelmonte:2007}) studied the effect of porosity and how it
changes over time, and they also used a Monte Carlo approach for it.

If one wishes to include particle properties in a traditional method which
solves the integral equations of coagulation (the Smoluchowski equation),
then one increases the dimensionality of the problem by 1 for each property
one adds. With only particle mass one has a distribution function $f(m,t)$
while adding two particle properties $p_1$ and $p_2$ means we get a
distribution function $f(m,p_1,p_2,t)$, making it a 4-dimensional problem.
Methods of this kind must treat the complete phase space spanned by
$(m,p_1,p_2,t)$. This is of course possible, but computationally it is
  a very challenging task (see Ossenkopf \citeyear{ossenkopf:1993}). In
contrast, a Monte Carlo method only sparsely samples phase space, and it
samples it only there where a significant portion of the total dust mass
is. A Monte Carlo method focuses its computational effort automatically
there where the action is. The drawback is that if one is interested in
knowing the distribution function there where only very little mass resides,
then the method is inaccurate. For instance, in a protoplanetary disk it
could very well be that most of the dust mass is locked up in big bodies
(larger than 1 meter) which are not observable, and only a promille of the
dust is in small grains, but these small grains determine the infrared
appearance of the disk because they have most of the solid surface area and
hence most of the opacity. In such a case a Monte Carlo method, by focusing
on where most of the mass is, will have a very bad statistics for those dust
grains that determine the appearance of the disk.  For such goals it is
better to use the traditional methods. But if we are interested in following
the evolution of the dominant portion of the dust, then Monte Carlo methods
naturally focus on the interesting parts of phase space.

\begin{figure*}
  \includegraphics[width=0.5\textwidth]{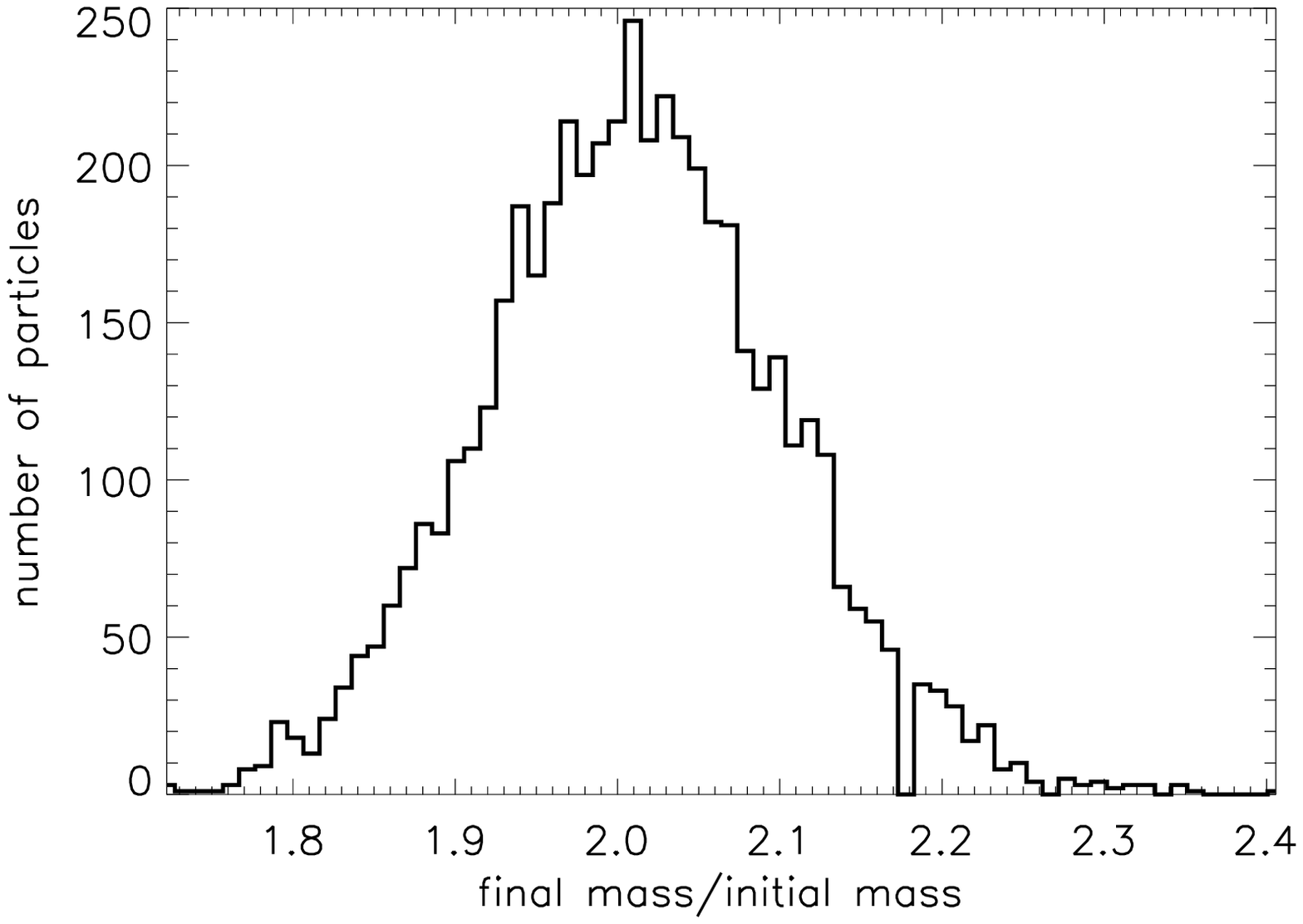}
  \includegraphics[width=0.5\textwidth]{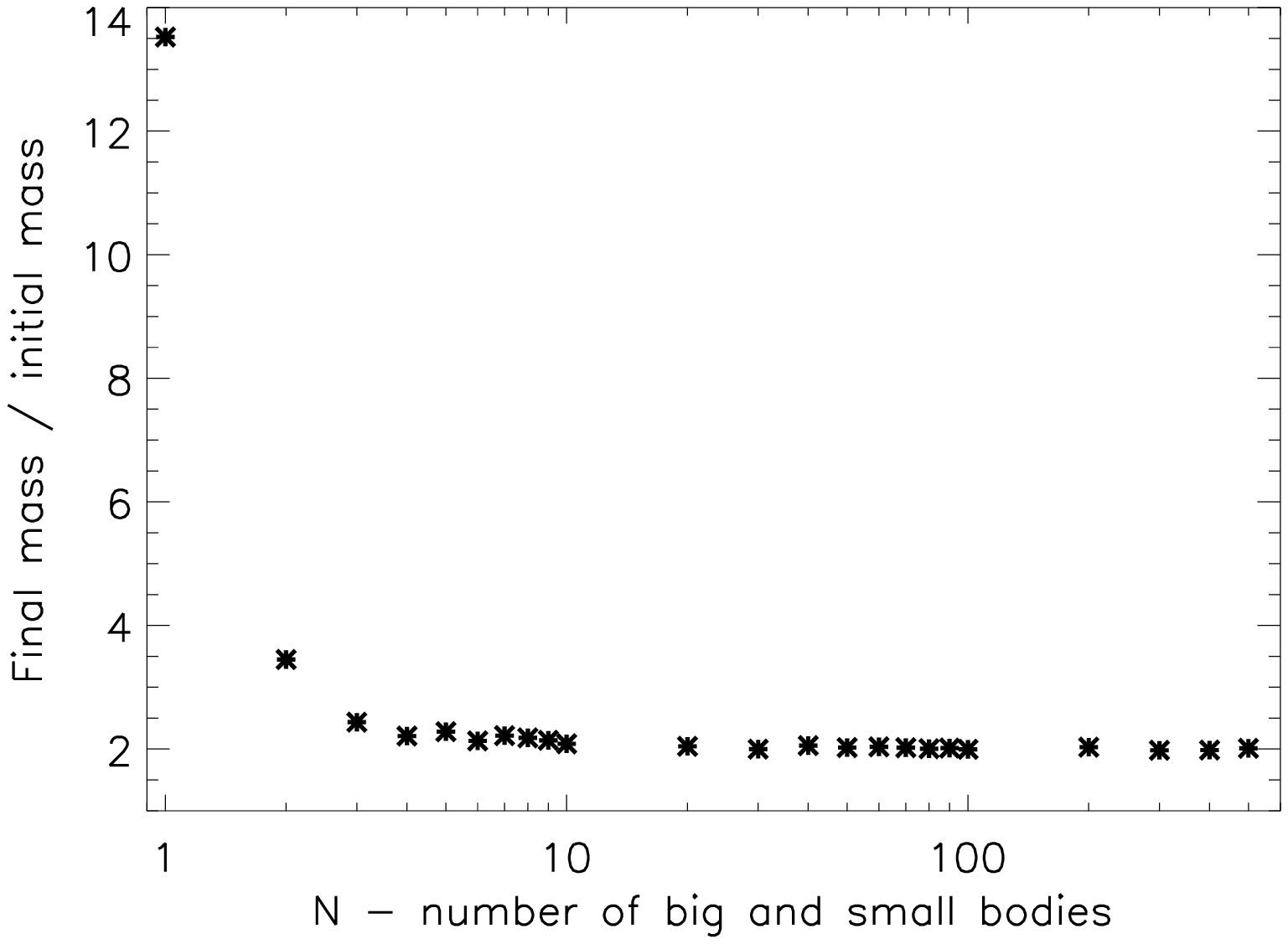}
\caption{\label{fig-exper-pairs} Results of the test problem with $N$ swarms
of small particles and $N$ swarms of big bodies, as discussed in Section \ref{subsec-cons-of-particle-number}. Left: histogram of the
final masses of the bodies relative to the initial mass of the big bodies,
for $N=500$. Right: average mass relative to the initial mass of the big bodies
as a function of $N$.}
\end{figure*}

\section{Discussion of the method}
\subsection{Conservation of particle number}
\label{subsec-cons-of-particle-number}
There are a few peculiarities of the method described here that may, at
first sight, appear inconsistent, but are statistically correct. For
instance if we return to our example of a swarm of tiny particles and a
swarm of boulders, i.e.\ $n=2$ with representative particle 1 being a micron
sized particle and representative particle 2 being a meter sized particle,
then we encounter an apparent paradox.  We again assume that collisions only
take place between 1 and 2, but not between 1 and 1 or 2 and 2. The chance
that representative particle 1 hits a meter size particle is much smaller
than the chance that representative particle 2 hits a micron size
particle. What will happen is that representative particle 2 will have very
many collision events with small micron size grains, and thereby slowly and
gradually grows bigger, while representative particle 1 will only have a
collision with big particle after a quite long time and immediately jumps to
that big size. While representative particle 2 grows in mass, the number of
big physical particles decreases in order to conserve mass. This may seem
wrong, because in reality the number of big boulders stays constant, and
these boulders simply grow by sweeping up the small dust. The solution to
this paradox is that the average time before representative particle 1 hits
a big ($k=2$) particle is of the same order as the time it takes for
representative particle 2 to grow to twice its mass by collecting small
particles. So, very roughly, by the time the big particle has doubled its
mass, and therefore the number of physical particles belonging to $k=2$ has
reduced by 50\%, the representative particle $1$ has turned into a big
particle, corresponding, statistically, to the other 50\% of big particles
that was missing. If we are a bit more precise, the statistics do not
  add up precisely in this way if we have only 1 swarm of small and 1 swarm
  of big bodies. If, however, one has N swarms of small and N swarms of big
  bodies, and again assume that only the big bodies can sweep up the smaller
  ones, then if $N\gg 1$ the statistics adds up perfectly: one finds that
  after all the growth has taken place, the average mass of the bodies is
  twice that of the original big bodies. In Figure~\ref{fig-exper-pairs} we do
  precisely this experiment, and the left panel shows that for $N=500$ the mass
  distribution of the big bodies averages to the right value, albeit with
  a spread of 10\% FWHM while in reality this spread should be 0. The
  right panel shows how the average final mass depends on $N$. For small
  $N$ the statistics clearly do not add up, but for large $N$ they do and
  produce the right value (final mass is twice initial mass of the big
  bodies). So statistically the number of big particles is restored to the
correct value, but there is then unfortunately still a large statistical
noise on it. The particle number is therefore not exactly conserved in our
method, but statistically it is.

\subsection{The number of representative particles}
\label{reppart}
It is obvious that for high number of representative particles $n$ we will
get better results than for low $n$. But there are two issues here.  First
of all, the higher $n$, the better the representative particles represent
the true physical distribution of particles. For problems that result in
wide size distributions this is all the more crucial. An inaccurate
representation of the true size distribution could lead to systematic
errors. But another reason for taking a high $n$ is simply because we want
our end-result to have as little as possible noise. If the result is too
noisy, then it is useless. Taking $n$ too big, however, makes the code slow
because more representative particles have to be followed, and for each of
these particles we must check for a larger number of possible collision
partners $k$. The problem scales therefore as $n^2$. If the expected size
distributions are not too wide, one can use an intermediately large number
of representative particles, say $\bar n$, for the simulation, but redo the
simulation $m$ times such that $n=m\bar n$, and average the results of all
$m$ simulations. This approach was also used by Ormel et
al.~(\citeyear{ormelmonte:2007}). This gives the same amount of noise on the
end-result, but scales as $\bar n^2m=n^2/m$, which is $m$ times faster than
the $n^2$ scaling. This works, however, only if the
coagulation/fragmentation kernel is not too sensitive to the exact
distribution of collision partners.

Interestingly, if the kernel is very insensitive to the exact distribution
of collision partners, then, in principle, one could run the model with only
a single representative particle $n=1$, because the collision partner of
representative particle $i$ could be equal to $k=i$. Of course, a single
representative particle means that we assume that all physical particles
have the same size, or in other words: that we have an infinitely narrow
size distribution. 

To decide about the sufficient number of representative particles, one has to compare the results of the MC code with the analytical solutions of the three test kernels (see Section \ref{section-test and results}). In a given time of the simulation the mean mass and the shape of the distribution function for all three test kernels must be followed accurately. It is especially important to reproduce the linear and product kernels accurately as the realistic kernels of dust particles are similar to these. 

Of course the progression from 'not sufficient' and 'sufficient' number of representative particles is smooth and in general the more representative particles we use, the more accurate the produced result will be. The sufficient number of representative particles ($\bar n$) as given in Section \ref{section-test and results}) are only suggestions, the error of the distribution functions were not quantified.

\subsection{Limitations of the method}
One of the fundamental limitations of the method described here is that we
assume $N\gg n$. We can model the growth of particles by coagulation in a
protoplanetary disk or in a cloud in a planetary atmosphere, but we can not
follow the growth to the point where individual large bodies start to
dominate their surroundings. For instance, if we wish to follow the growth
of dust in a protoplanetary disk all the way to small planets, then the
method breaks down, because $N$ is then no longer much bigger than $n$, and
interactions among representative particles become likely. Also, for the
same reason, run-away growth problems such as electrostatic gelation (Mokler
\citeyear{mokler:2007}) cannot be modeled with this method. 

Another limitation is encountered when modeling problems with strong growth
and fragmentation happening at the same time. This leads to very wide size
distributions, and the typical interval between events is then dominated by
the smallest particles, whereas we may be interested primarily in the growth
of the biggest particles. In such a situation a semi-steady-state can be
reached in which particles coagulate and fragment thousands of times over
the life time of a disk. The Monte Carlo method has to follow each of these
thousands of cycles of growth and destruction, which makes the problem very
``stiff''.  Methods using the integral form of the equations, i.e.\ the
Smoluchowski equation, can be programmed using implicit integration in time
so that time steps can be taken which are much larger than the typical time
scale of one growth-fragmentation cycle without loss of accuracy (Brauer et
al.~\citeyear{brauer:2008}). This is not possible with a Monte Carlo method.

\section{Standard tests and results}
\label{section-test and results}
In this section we test our coagulation model with kernels that have
analytical solutions. Furthermore we show the first results of applying this
model to protoplanetary disks introducing Brownian motion and turbulence
induced relative velocities as well as a new property of dust particles
namely the porosity (or enlargement factor, see Ormel et al
\citeyear{ormelmonte:2007}), and a simple fragmentation model.

\begin{figure}
  \centering
  \includegraphics[width=0.5\textwidth]{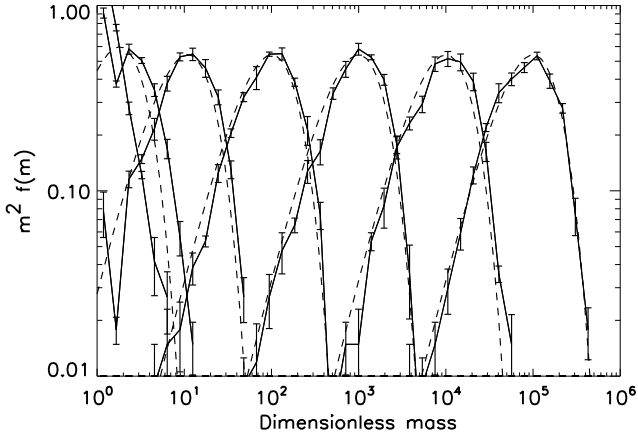}
  \caption{Test against the constant kernel ($K_{i,j}=1$). The particles were binned and the distribution function was produced at dimensionless times $t=0,10^0,10^1,10^2,10^3,10^4,10^5$. The dashed lines show the analytical solution. This run was produced by simulating 200 representative particles five times and producing the average of these. In this case $dm_{max}$ is 0.1.}
  \label{const}
\end{figure}

To follow dust coagulation and fragmentation, one has to follow the time
evolution of the particle distribution function at a given location in the
disk ($f(\bf{y},t)$), where $\bf{y}$ contains the modeled properties of the
dust grains, in our case these will be the mass ($m$) and the enlargement
factor ($\Psi$), $f(m,\Psi,t)$.

In most of the coagulation models so far the only used dust-property was the
particle mass. Then one can use the so called Smoluchowski equation
(\cite{smoluchowski:1916}) to describe the time-evolution of $f(m)$:
\begin{multline}
\frac{\partial f(m)}{\partial t}=-f(m)\int dm' K(m,m')f(m') + \\
 \frac{1}{2}\int dm' K(m',m-m')f(m')f(m-m').
\end{multline}
The first term on the right hand side represents the loss of dust in the
mass bin $m$ by coagulation of a particle of mass $m$ with a particle of
mass $m'$. The second term represents the gain of dust matter in the mass
bin $m$ by coagulation of two grains of mass $m'$ and $m - m'$. $K$ is the
coagulation kernel, it can be written as
\begin{equation}
K(m_1,m_2)=\sigma_c(m_1,m_2)\times \Delta v (m_1,m_2),
\end{equation} 
the product of the the cross-section of two particles and their relative
velocity. We consider all the three kernels for which there exist
  analytical solutions: The constant kernel ($K_{i,j}=1$), the linear
kernel ($K_{i,j}=m_i+m_j$) and the product kernel
  ($K_{i,j}=m_i\times m_j$). The analytical solutions are described e.g.
in \cite{ohtsuki:1990} and Wetherill (\citeyear{wetherill:1990}).

\begin{figure}
  \centering
  \includegraphics[width=0.5\textwidth]{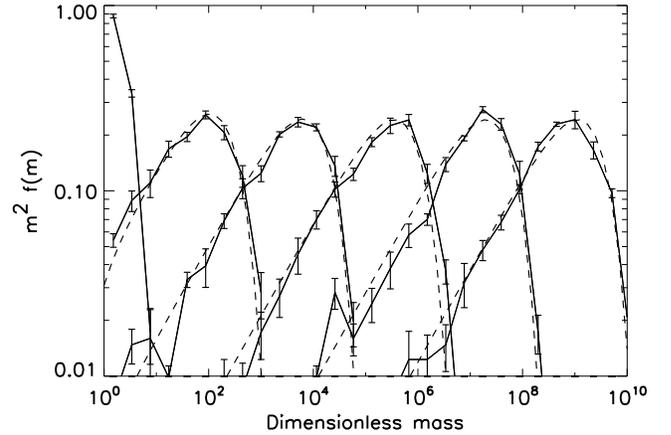}
  \caption{Test against the linear kernel ($K_{i,j}=m_i+m_j$). The particles were binned and the distribution function was produced at dimensionless times $t=0, 4, 8, 12, 16, 20$. The dashed lines show the analytical solution. This run was also produced by simulating 200 representative particles five times and producing the average of these. In this case $dm_{max}$ is 0.1.}
  \label{lin}
\end{figure}

We test our method against these three kernels, leaving the
enlargement factor unchanged, always unity. Further important properties
of the dust particles, such as material density and volume density, are also
always unity. The (dimensionless) time evolution of the swarms is followed
and at given times the particles are binned by mass so that we can produce
$f(m)$. On Figures \ref{const} and \ref{lin} the y axis shows $f(m)\times
m^2$, the mass density per bin. The analytical solutions, taken from
\cite{ohtsuki:1990} and Wetherill
  (\citeyear{wetherill:1990}), are overplotted with dashed
line. The number of particles were chosen to be $\bar n=200$, $m=5$, so
altogether 1000 representative particles were used in the model
except for the product kernel where more representative particles
  were used to achieve better results.

\begin{figure}
  \centering
  \includegraphics[width=0.5\textwidth]{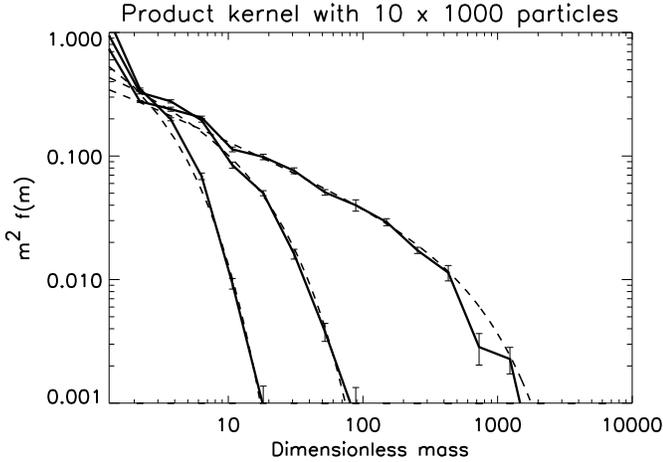}
  \caption{Test against the product kernel ($K_{i,j}=m_i \times m_j$). The particles were binned and the distribution function was produced at dimensionless times $t=0.4,0.7, 0.95$. The dashed lines show the analytical solution. This run was produced by simulating 1000 representative particles ten times and producing the average of these. In this case $dm_{max}$ is set to be  0.05.}
  \label{prod}
\end{figure}

In the case of the constant kernel (Figure~\ref{const}), we started our
simulation with MRN size distribution ($n(a)\propto a^{-3.5}$), the results
were saved at $t=0,10^0,10^1,10^2,10^3,10^4,10^5$. It is interesting to note
that this kernel is not sensitive to the initial size distribution. As the
system evolves, it forgets the initial conditions. Another interesting
property of this kernel that our model can reproduce the analytical solution
even with very limited number of representative particles (even for $\bar
n=5$!) but of course with higher noise. It is possible to use only one
representative particle, which means that the representative particle
collides with particles from its own swarm which basically results in pure
CCA growth (Cluster-Cluster Aggregation). Interestingly, the mean mass of the distribution function is
followed correctly but the shape of the function changes, additional spikes
appear on it. 

The linear kernel is known to be more problematic because the mean mass of
the particles grows exponentially with time. Our model, however reproduces
this kernel very well, too, as it can be seen in Figure \ref{lin}. The results
were saved at $t=0, 4, 8, 12, 16, 20$. We note that using low number of
representative particles with this kernel also works relatively well, the
minimum number of swarms needed to reproduce the exponential time evolution
of the mean mass is $\bar n \approx 100$.  This is larger than for the
constant kernel. It shows that for the linear kernel collisions between
particles of unequal mass are contributing significantly to the growth,
whereas for the constant kernel the growth is dominated by collisions
between roughly equal size particles. Using $\bar n \ll 100$ results in
distorted distribution function: neither the mean mass nor the actual shape
of the distribution function is correct.

The product kernel is the hardest to reproduce. The peculiarity of
  this kernel is the following: Using dimensionless units, a 'run-away'
  particle is produced around $t=1$, which collects all the other particles
  present in the simulation (Wetherill \citeyear{wetherill:1990}). The
  difficulty arises in our Monte Carlo code when the mass of the
  representative 'run-away' particle reaches the mass of its swarm. In other
  words, the number of physical particles belonging to the
  representative 'run-away' particle is close to unity. In this case the
  original assumption of our method (we only need to consider collisions
  between a representative particle and a physical particle) is
  not valid anymore. However, as Figure \ref{prod} shows, we can relatively
  well reproduce this kernel before $t=1$. In the case of this kernel, we
  need approximately $\bar n \approx 500$ representative particles to
  correctly reproduce it.

The required CPU time for these test cases is very low, some seconds
  only.

We conclude that our Monte Carlo method reproduces the constant and linear test kernels without any problem even with low number of representative particles. On the other hand the method has difficulties with the product kernel, but before the formation of the 'run-away' particle, we can reproduce the kernel. The relatively low number of representative particles needed to sufficiently reproduce the test kernels is very important for future applications where whole disk simulations will be done
and there will likely be regions containing low numbers of particles.

\section{Applications to protoplanetary disks}
We use the Monte Carlo code to follow the coagulation and fragmentation of
dust particles in the midplane of a protoplanetary disk at 1 AU from the
central star. Our disk model is identical with the one used by
\cite{brauer:2007}. We proceed step by step. First relative velocities
induced by Brownian motion and turbulence without the effects of porosity
are included (Sec.~\ref{relv}).

The next step is to include a fragmentation model (Sec.~\ref{fragmod}). 

In the final step porosity is included (Sec.~\ref{poro}). We use the
porosity model described in Ormel et al.~(\citeyear{ormelmonte:2007}). At this point we compare and check again our code with Ormel et al.~(\citeyear{ormelmonte:2007}) using their input parameters but not including the rain out of particles.

\subsection{Relative velocities}
\label{relv}
We include two processes in calculating the relative velocities: Brownian
motion and turbulence. 

Brownian motion strongly depends on the mass of the two colliding
particles. The smaller their masses are, the more they can be influenced by
the random collisions with the gas molecules/atoms. One can calculate an
average velocity given by
\begin{equation}
 \Delta v_B (m_1,m_2) = \sqrt \frac{8kT(m_1+m_2)}{\pi m_1 m_2}.
\end{equation}
For micron sized particles, relative velocity can be in the order of
magnitude of 1 cm/s, but for cm sized particles this value drops to
$10^{-7}$ cm/s. If growth is only governed by Brownian motion, it leads to
very slow coagulation, a narrow size distribution and fluffy dust particles,
so called cluster-cluster aggregates (CCA).

The gas in the circumstellar disk is turbulent, thus the dust particles
experience acceleration from eddies with different sizes and turnover
times. This process is very complex, but Ormel and Cuzzi
(\citeyear{ormelcuzzi:2007}) provided limiting closed-form
expressions for average relative turbulent velocities between two dust
particles. Their results are also valid for particles with high Stokes
numbers. They distinguished three regimes: a.) the stopping times of both
dust particles are smaller than the smallest eddy-turnover time ($t_1,t_2 <
t_{\eta}$, tightly coupled particles); b.) the stopping time is between the
smallest and largest turnover time ($t_{\eta} \le t_1 \le t_L$, intermediate
regime); c.) the stopping time is bigger than the largest turnover time
($t_1 > t_L$, heavy particles). For details see Ormel and Cuzzi
(\citeyear{ormelcuzzi:2007}). We used $ \alpha = 10^{-3}$ for the turbulence
parameter.

\begin{figure}
  \centering
  \includegraphics[width=0.5\textwidth]{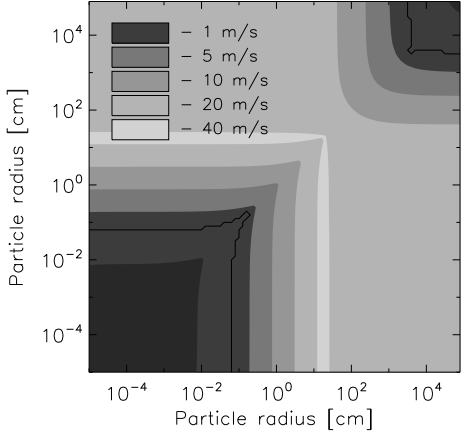}
  \caption{The relative velocity caused by Brownian motion and turbulence
    for different sized particles. The black line shows the fragmentation
    barrier. Collision events situated between these two lines result in
    fragmentation if porosity is not included. Physical parameters of the
    disk: the distance from the central star is 1 AU, temperature is 200 K,
    the density of the gas is $8.73\times 10^{12}$ particle/cm$^3$, and the
    turbulent parameter, $\alpha=10^{-3}$. Parameters of the dust: monomer
    radius is $a_0 =0.4 \mu$m, material density is $\rho = 1.6$ g/cm$^3$.}
  \label{relcont}
\end{figure}

To illustrate the relative velocity of dust particles without the effects of
porosity, we provide Figure \ref{relcont}. This contour plot includes Brownian
motion and turbulent relative velocities. The Brownian motion is negligible
for particles bigger than $10^{-2}$ cm.

\subsection{Fragmentation model} 
\label{fragmod}
The collision energy of the particles is
\begin{equation}
  E=\frac{1}{2} \frac{m_1 m_2}{m_1 + m_2} \Delta v^2 = \frac{1}{2} \mu \Delta v^2,
\end{equation}
where $\mu$ is the reduced mass. We need to define some quantities of the
dust particles. $E_{roll}$ is the rolling energy of two monomers. For
monomers of the same size it is given by (Dominik \& Tielens
\citeyear{dominiktielens:1997}; Blum \& Wurm \citeyear{blumwurm:2000})
\begin {equation}
  E_{roll}=\frac{1}{2}\pi a_0 F_{roll},
  \label{eq:Eroll}
\end{equation}
where $a_0$ is the monomer radius, $F_{roll}$ is the rolling force measured
by Heim et al.~(\citeyear{heim:1999}). Its value is $F_{roll}=(8.5 \pm 1.6)
\times 10^{-5}$ dyn for SiO$_2$ spheres.

The fragmentation energy is then defined as follows:
\begin{equation}
E_{frag} = N_c \times E_{break} \simeq 3 N \times E_{roll},
\label{eq:Efrag}
\end{equation}
where $N_c$ is the total number of contact surfaces between monomers (for
simplicity it is taken to be 3N, where N is the number of monomers in the
particle), $E_{break}$ is the energy needed to break the bond between two
monomers (its order of magnitude is similar to $E_{roll}$ for these parameters).

If the collision energy of two particles is higher than the corresponding
fragmentation energy, then the aggregate is destroyed and monomers are
produced. Note that although assuming a complete destruction of the
  collided dust particles, we are interested in the critical energy where
  the first fragmentation event happens. This is the reason why the
  fragmentation energy is assumed to be lower than the energy needed for
  catastrophic fragmentation. It is a simplification of the model to
  assume that the debris particles will be monomers. This is a very
simplified fragmentation model used previously by Dullemond \& Dominik
(\citeyear{duldom:2005}). A more realistic model would be the one used by
\cite{brauer:2007}.

We show the fragmentation barrier in Figure \ref{relcont} with black lines. If
collision happens in the regime between these two lines, that results in
fragmentation.

\subsubsection{Results}
A simulation was made including these effects in a specified location of the
disk. We choose the location to be 1 AU distance from the central solar type
star. Using the disk model of \cite{brauer:2007}, the temperature at this
distance is approximately 200 K, the density of the gas is $8.73\times
10^{12}$ cm$^{-3}$, the gas-to-dust ratio is 100 and we choose the
turbulent parameter to be $\alpha=10^{-3}$, the Reynolds number is
$Re=10^{8}$ (based on Ormel \& Cuzzi \citeyear{ormelcuzzi:2007}). The dust
monomers have the following properties: the monomer radius is $a_0 =0.4$ $\mu$m, material density is $\rho = 1.6$ g/cm$^3$. With the used parameters
the fragmentation velocity is $\Delta v_{frag} \approx 8$ m/s, though it is
somewhat larger for equal sized agglomerates. It is important to note that
this value is very sensitive to the monomer radius ($a_0$) and material
density ($\rho$), because smaller/lighter monomers mean more contact
surfaces (higher N for the same mass) and therefore higher fragmentation
energy.

\begin{figure}
\centering
\includegraphics[width=0.5\textwidth]{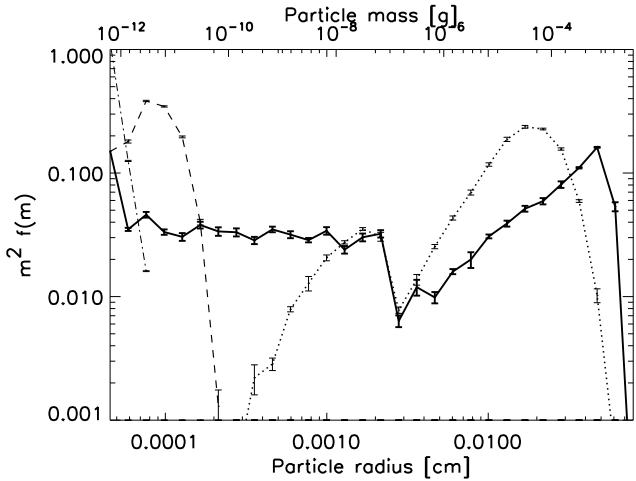}
\caption{The evolution of dust particles including the effects of Brownian
  motion and turbulence. Porosity is not included in this model. The
  particle distribution is saved after $t=3\times 10^0$ years - dash-dot
  line, $3\times 10^1$ years - dashed line, $3\times 10^2$ years - dotted
  line, and $3\times 10^3$ years - continuous line.}
\label{res_nocomp}
\end{figure}

Using these input parameters we simulated the evolution of the dust
particles for $3\times 10^3$ years so that we reach an equilibrium between
coagulation and fragmentation. Figure~\ref{res_nocomp} shows the resulting
normalized size distributions in times after $t=3\times 10^0$, $3\times
10^1$, $3\times 10^2$ and $3\times 10^3$ years. We used $\bar n=100$
particles averaging over $m=100$ times ($10^4$ particles
  altogether). The required CPU time to perform this simulation is 1.5 hours
  approximately. $dm_{max}$ is set to be 0.001 from now on in every simulation. We would like to note that giving $dm_{max}$ (Section \ref{subsec-acceleration}) a higher value would decrease the CPU time.

One can see that coagulation happens due to Brownian motion in the beginning
of the simulation (until $3\times 10^1$ years) but after that turbulence
takes over and the first fragmentation event happens after roughly $10^3$
years. After this event the "recycled" monomers start to grow again, but as
we see in Figure \ref{relcont}, particles can not reach bigger sizes than 0.07 cm.

We would like to draw attention to the sudden decrease of particles
  around 0.002 cm in Figure~\ref{res_nocomp}. This is the result of the
  turbulent relative velocity model used here (discussed in
  Sect. \ref{relv}). At this point the particles leave the 'tightly coupled
  particles' regime and enter the 'intermediate' regime. But the transition
  in relative velocity between these regimes is not smooth, there is a jump
  in relative velocity from $\sim$20 cm/s to $\sim$60 cm/s. As a result,
  particles coagulate suddenly faster and leave this part of the size
  distribution rapidly. Similar 'valleys' can be seen in the following
  figures with porosity, but the feature is less distinct as the stopping
  times can be different for particles with same mass.


 
\subsection{Porosity}
\label{poro}

 \begin{figure}
\centering
\includegraphics[width=0.5\textwidth]{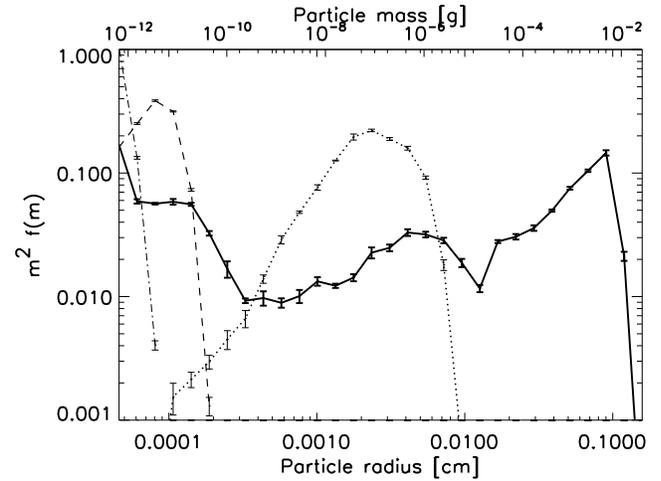}
\caption{The evolution of dust particles including the effects of Brownian
  motion and turbulence. Porosity is included in this model! The x axis
  shows the compact radius. The particle distribution is saved after
  $t=3\times 10^0$ years - dash-dot line, $3\times 10^1$ years - dashed
  line, $3\times 10^2$ years - dotted line, and $3\times 10^3$ years -
  continuous line. Note that the scaling of the x axis is different
    from Figure~\ref{res_nocomp}.}
\label{res_comp}
\end{figure} 

To be able to quantitatively discuss the effect of porosity, we have to
define the enlargement parameter following the discussion of Ormel et
al.~(\citeyear{ormelmonte:2007}). If $V$ is the extended volume of the grain
and $V^*$ is the compact volume, than one can define the enlargement
parameter ($\Psi$) as

\begin{equation}
\Psi = \frac{V}{V^*}.
\end{equation}  
Compact volume is the volume occupied by the monomers not taking into
account the free space between the monomer spheres. One can think of it as
melting all the monomers into a single sphere, the volume of this sphere is
the compact volume. We use compact radius later on, which is the radius of
this sphere. In the previous section the mass/volume ratio was constant for
the particles. Therefore we could automatically calculate the mass of the
particle if the radius was known or vice-versa. But from now on a particle
with given mass $m$ can have a wide range of effective radii depending on
its enlargement parameter.

It is essential to know how the enlargement parameter changes upon
collisions. We have to refine our fragmentation model and introduce two more
regimes regarding to collision energy. We use the model of Ormel et 
al.~(\citeyear{ormelmonte:2007}) and we only summarize their model here.

The first regime is the low collision energy regime, where the collision
energy is smaller than the restructuring energy ($E < E_{restr}$, where
$E_{restr} = 5 E_{roll}$), meaning that the particles stick where they meet,
the internal structure of the grain does not change.

The recipe for the resulting enlargement factor after the collision of two
particles assuming that $m_1 > m_2$ then is
\begin{equation}
\Psi = \langle \Psi \rangle _m \left(1+\frac{m_2 \Psi_2}{m_1 \Psi_1}  \right)^{\frac{3}{2}\delta_{CCA}-1} + \Psi _{add},
\end{equation}   
where  $\langle \Psi \rangle _m$ is the mass averaged enlargement factor of
the colliding particles:
\begin{equation}
\langle \Psi \rangle _m = \frac{m_1 \Psi_1 + m_2 \Psi_2}{m_1 + m_2}.  
\end{equation}  
Furthermore $\delta_{CCA}$ is the CCA-characteristic exponent calculated by
detailed numerical studies such as Paszun \& Dominik (\citeyear{paszun:2006})
($\delta_{CCA}=0.95$). $\Psi _{add}$ is a necessary additional factor for
the enlargement factor (for details see Ormel et
al.~\citeyear{ormelmonte:2007}): 
\begin{equation}
  \Psi _{add} = \frac{m_2}{m_1} \Psi_1 \exp \left[ \frac{-\mu}{10 m_0} \right],
\end{equation}
where $m_0$ is the monomer mass.

The second regime is the regime of compaction. The internal structure of the
monomers inside the particle changes, this causes a decreasing porous
volume. If the collision energy $E_{restr} \le E \le E_{frag}$, we talk
about compaction. In this case the porosity after the collision becomes
\begin{equation}
 \Psi = (1-f_C)(\langle \Psi \rangle _m - 1 ) +1,
 \label{eq:comp}
\end{equation}
where $f_C=E/(N E_{roll})=-\Delta V / V$ is the relative compaction. One can
see that $f_C$ has to be smaller than unity otherwise  $\Psi$ in Eq. \ref{eq:comp}
becomes less than unity. But it can theoretically happen that $E> N
E_{roll}$. In this case, as long as the total collision energy remains below
the fragmentation threshold, we assume that after compaction this excess
energy goes back into the kinetic energy of the two colliding aggregates.
The two aggregates therefore compactify and bounce, without exchanging mass
or being destroyed. Bouncing is therefore included in this model, albeit in
a crude way.

The third regime is fragmentation as it was discussed in the previous
section ($E > E_{frag}$). We use the same fragmentation model as before so
the result of a fragmenting collision are monomers.

\subsubsection{Results}
\label{subsec:res_por}
We performed a simulation with exactly the same initial conditions as in the
last section but we included the porosity as an additional dust property in
the model. The result can be seen in Figure~\ref{res_comp} (the
  required CPU time here is also 1.5 hours). One can immediately see that
including porosity increases the maximum particle mass by two orders of
magnitude (five times larger particles in radius). This was already expected
based on the work of Ormel et al.~(\citeyear{ormelmonte:2007}), although due
to rain out of bigger particles, they did not simulate particles bigger than
0.1 cm.

\begin{figure}
\centering
\includegraphics[width=0.5\textwidth]{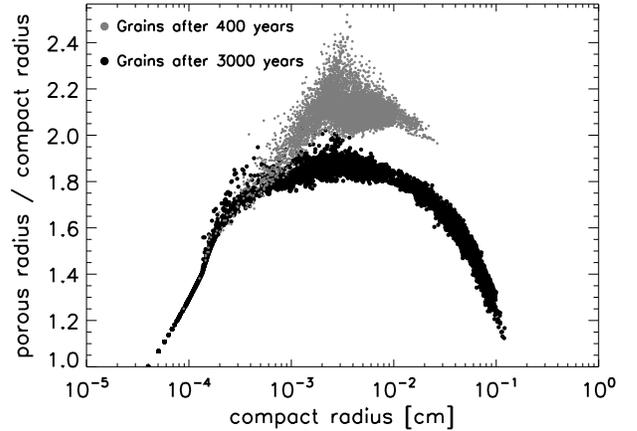}
\caption{This figure shows the radial enlargement of the dust aggregates after 400 and 3000 years. The x axis is the compact radii of the particles, the y axis is the ratio between the compact and the porous radius, this quantity is basically equal to $\Psi ^{\frac{1}{3}} $.}
\label{res_fluffy}
\end{figure}

We provide Figure~\ref{res_fluffy} to give an impression how the
porosity of the agglomerates change during the simulation. The x axis is the
compact radius of the particles, the y axis is the ratio between the compact
and the porous radii. This quantity is basically equal to $\Psi
^{\frac{1}{3}}$. Fractal growth is important for small particles creating
fluffy agglomerates (until $10^{-3}-10^{-2}$ cm approximately), after this
point the relative velocities become high enough so compactness becomes
important. Before the particles reach a fully compacted stage they fragment,
become monomers and a new cycle of growth starts. It is important to note that the porosity of the aggregates before the first fragmentation event is usually higher than the porosity values after equilibrium is reached. This can be seen in Figure~\ref{res_fluffy} (grains after 400 years and 3000 years). The reason is that before the first fragmentation event, particles involved in collisions are typically equal sized so these particles produce fluffy structures. However, when the distribution function relaxes in equilibrium, there are collisions between smaller and bigger aggregates as well which results in somewhat compacted aggregates.

\subsubsection{Model comparison with Ormel et 
al.~(\citeyear{ormelmonte:2007})}
We compare our Monte Carlo code with the one developed by Ormel et
al.~(\citeyear{ormelmonte:2007}). They use the Minimum Mass Solar Nebula
disk model (MMSN) and somewhat different dust-parameters which we changed
accordingly (distance from the central star = 1 AU, temperature = 280 K,
density of the gas is 8.5 $\times$ 10$^{-10}$ g/cm$^3$, gas to dust ratio =
240, $\alpha$ = 10$^{-4}$; monomer radius = 0.1 $\mu$m, monomer density = 3
g/cm$^3$, surface energy density of the monomers = 25 ergs/cm$^{-2}$).

They follow particle coagulation at one pressure scale height above the
midplane of the disk. Because of this if the particles reach a critical
stopping time ($\tau_{rain}=\alpha / \Omega$, where $\Omega$ is the Kepler
frequency), the particles rain out meaning that these particles leave the
volume of the simulation, the distribution function of the dust particles is
collapsing as it can be seen in their figures (Figure 10 and 11 in Ormel et
al.~(\citeyear{ormelmonte:2007})).

We do not include this effect in our model but we stop the simulation at the
first rain out event and compare our distribution functions until this
point. We use $10^4$ representative particles ($100\times 100$)
  during the simulation.

\begin{figure}
\centering
\includegraphics[width=0.5\textwidth]{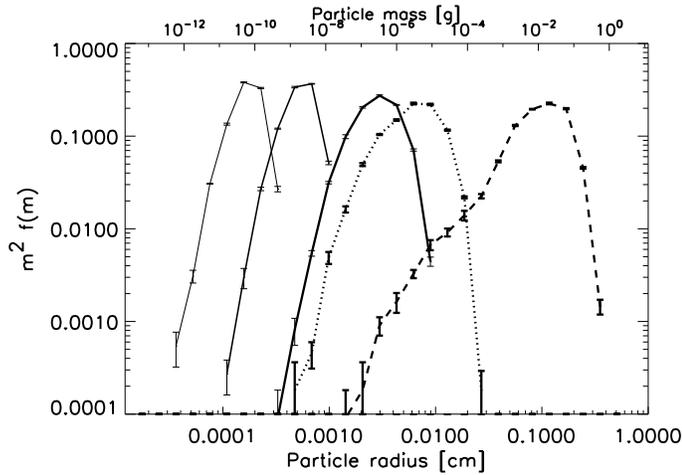}
\caption{Distribution functions obtained by using Ormel et
  al.~(\citeyear{ormelmonte:2007}) input parameters. The continuous lines
  show the distribution functions at t = 10 years (thin line), 100 years
  (thicker line), 1000 years (thickest line). The dotted line shows the
  distribution function at the time of the first compaction event (t = 1510
  years), the dashed line shows the distribution function at the first rain
  out event (t = 2900 years).}
\label{ormel}
\end{figure}

This can be seen at Figure~\ref{ormel}. The reader is advised to examine this
figure together with Figure 10. c. from Ormel et
al.~(\citeyear{ormelmonte:2007}) because this is the figure we reproduced
here. Furthermore we would like to point out that the scale of the y axis is
different in the two figures. Our figure shows two orders of magnitude from
the normalized distribution functions whereas their figure covers more than
10 orders of magnitude from the real distribution function.

Keeping these in mind, one can compare the results of the two Monte Carlo
codes.

The continuous lines at Figure~\ref{ormel} in this paper show the distribution
functions at t = 10 years (thin line), 100 years (thicker line), 1000 years
(thickest line). The dotted line shows the distribution function at the time
of the first compaction event (t = 1510 years), the dashed line shows the
distribution function at the first rain out event (t = 2900 years). The same
notation is used by Ormel et al.~(\citeyear{ormelmonte:2007}) at Figure 10. c.

We compared the position of the peaks of the distribution functions and the approximate shape of the curves. We can conclude that our code reproduces the results of Ormel et al.~(\citeyear{ormelmonte:2007}) very well.  

The required CPU time to perform this simulation is only 10
  minutes. One might ask why the CPU time is almost ten times smaller now?
  Why do the previous simulations, which used the same number of
  representative particles ($10^4$) and simulated approximately the same
  time interval (3000 years), take so long? The required CPU time does not
  scale linearly with the used number of particles. It scales linearly with
  the number of collisions simulated. The difference between this run and
  the previous two simulations is fragmentation. In the simulations of Ormel
  et al.~(\citeyear{ormelmonte:2007}) no fragmentation is happening because
  the growth timescales are longer. Using our initial parameters, the first
  fragmentation event happens around 1000 years, the number of small
  particles are never completely depleted after this time. As the small
  particles thereafter are always present, the number of collisions will be
  much higher than before.

Also note that the porosities of these particles would be smaller if
  the model of Ormel et al.~(\citeyear{ormelmonte:2007}) included
  fragmentation (for the reason see Sect. \ref{subsec:res_por}).

\subsection{Monomer size distribution}

\begin{figure}
\centering
\includegraphics[width=0.5\textwidth]{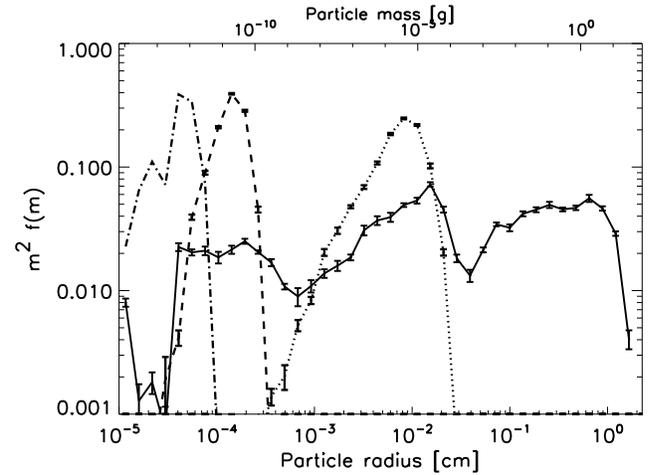}
\caption{The evolution of dust particles including the effects of Brownian
  motion and turbulence, porosity and using two different monomer sizes
  ($a_1=0.1 \mu$m and $a_2=0.4 \mu$m). The particle distribution is saved
  after $t=3\times 10^0$ years - dash-dot line, $3\times 10^1$ years -
  dashed line, $3\times 10^2$ years - dotted line, and $3\times 10^3$ years
  - continuous line.}
\label{2mon}
\end{figure}  

An interesting question which can easily be answered with our method
  is: How the mixture of different sized monomers change the maximum
  agglomerate size which can be reached? As we can see from Equations
  \ref{eq:Eroll} and \ref{eq:Efrag}, the rolling energy is lower for smaller
  monomers and of course the number of monomers in an agglomerate is much
  higher if the same agglomerate is built up of lighter monomers. This would
  mean higher fragmentation energy and one would expect that the particles
  would be harder to fragment resulting in bigger grains.

We performed a simplified simulation to be able to answer this
  question. Only two different monomer sizes are considered here, $a_1=0.1
  \mu$m and $a_2=0.4 \mu$m assuming that half of the mass (or representative
  particles) belongs to the small monomers, the other half belongs to the
  big monomers.

One problem arises here with the rolling energy. The rolling energy
  changes with monomer size, and as our method cannot follow exactly the
  number of contacts in an aggregate and what kind of monomers are
  connected, we are forced to use an averaged rolling energy. One has to
  carefully consider what the average rolling energy should be. In our case,
  the big monomer is 64 times heavier than the small monomer. Let's assume
  that 50\% of the mass of an aggregate is built up from small monomers; on
  the other hand, if we compare the number of different monomers, the small
  monomers will be 64 times more numerous than the big ones. This means that
  the contribution of small monomers in the average rolling energy ($\bar
  E_{roll}$) should be higher. This can be achieved by using the following
  weighting:
\begin{equation}
\bar E_{roll} = \frac{a_1}{a_1+a_2} E_{roll_2}+ \frac{a_2}{a_1+a_2} E_{roll_1},
\end{equation}
where $E_{roll_1}$ is the rolling energy between monomers with
  radius $a_1$, $E_{roll_2}$ is the rolling energy between monomers with
  radius $a_2$.

As we can see in Figure \ref{2mon}, the maximum aggregate sizes
  reached are approximately an order of magnitude higher than on Figure
  \ref{res_comp} as it was predicted earlier in this Section.

\section{Conclusions and outlook}
We have shown that our representative particle method for aggregation of
particles in astrophysical settings works well for standard kernels. It has
the usual advantages of Monte Carlo methods that one can add particle
properties easily and without loss of computational speed. Moreover, it
naturally conserves the number of computational elements, so there is no
need to ``add'' or ``remove'' particles. Each representative particle
represents a fixed portion of the total mass of solids. 

Our method may have various possible interesting extensions and
applications. Here we speculate on a few of these. For instance, the fact
that each representative particle corresponds to a fixed amount of solid
mass makes the method ideal for implementation into spatially resolved
models such as hydrodynamic simulations of planetary atmospheres or
protoplanetary disks. We can then follow the exact motion of each
representative particle through the possibly turbulent environment, and
thereby automatically treat the stochastic nature and deviation from a
Boltzmann distribution of the motion of particles with stopping times of the
same order as the turbulent eddy turn-over time. It is necessesary,
  however, to assure that a sufficiently large number of representative
  particles is present in each grid cell of the hydrodynamic simulation.
  For large scale hydrodynamic simulations this may lead to a very large
  computational demand for the coagulation computation, as well as for
  tracking the exact motion of these particles. If strong clumping of the
  particles happens, however, much of the ``action'' anyway happens in these
  ``clumps'', and it may then not be too critical that other grid cells are
  not sufficiently populated by representative particles. This, however, is
  something that has to be experimented.

Our representative particle
method can in principle also be used to model the sublimation and
condensation of dust grains. If a particle sublimates then the
representative particle becomes simply an atom or molecule of the vapor of
this process. It will then follow the gas motion until the temperature
becomes low enough that it can condense again. Other representative
particles which are still in the solid phase may represent physical
particles that can act as a condensation nucleus. Finally, in our method the
properties of the particle can not only change due to collisions, but we can
easily implement other environmental factors in the alteration of particle
properties.

There are two main drawbacks of the method. First, it only works for large
particle numbers, i.e.\ it cannot treat problems in which individual
particles start dominating their immediate environment. Ormel's method and
its expected extension do not have this problem. Secondly, the method
cannot be accelerated using implicit integration, while Brauer's method can.

All in all we believe that this method may have interesting applications 
in the field of dust aggregation and droplet coagulation in protoplanetary
disks and planetary atmospheres.

\begin{acknowledgement}
  We wish to thank Frithjof Brauer, Anders Johansen, Patrick Glaschke,
  Thomas Henning, J\"urgen Blum, Carsten G\"uttler, Carsten Dominik and
  Dominik Paszun for useful comments. We also thank the anonymous referee
  for very useful comments that helped to improve the paper and for pointing
  us to a problem in our presentation of the solution to the particle number
  conservation paradox.
\end{acknowledgement}


\end{document}